\documentclass[runningheads]{llncs}
\usepackage{graphicx}
\usepackage[linesnumbered,ruled]{algorithm2e}
\usepackage{soul}
\usepackage{color,xcolor}
\usepackage{amsfonts}
\usepackage{hyperref}
\usepackage{enumitem}
\usepackage{amsmath}
\usepackage{threeparttable}
\usepackage{amsmath}
\usepackage{amsfonts}
\usepackage{makecell}
\usepackage{caption}
\usepackage{booktabs}
\usepackage{array}
\usepackage{cases}
\usepackage{threeparttable}
\usepackage{multicol}
\usepackage{enumitem}
\usepackage{marvosym}
\hypersetup
{
colorlinks=true,
linkcolor=blue,
filecolor=blue,
urlcolor=blue,
}
\begin{document}
\title{BigFUSE: Global Context-Aware Image Fusion in Dual-View Light-Sheet Fluorescence Microscopy with Image Formation Prior}
\titlerunning{BigFUSE}
%
\author{Yu Liu\inst{1}\orcidID{0000-0003-2281-6791} \and
Gesine Müller\inst{2} \and
Nassir Navab\inst{1, 4}\orcidID{0000-0002-6032-5611} \and
Carsten Marr\inst{5}\orcidID{0000-0003-2154-4552} \and
Jan Huisken\inst{2,3}\orcidID{0000-0001-7250-3756}\textsuperscript{\Letter} \and
Tingying Peng\inst{6}\orcidID{0000-0002-7881-1749}\textsuperscript{\Letter}}
\authorrunning{Y. Liu et al.}

\institute{Technical University of Munich, Munich, Germany \and
Georg-August-University Göttingen, Göttingen, Germany \and Cluster of Excellence "Multiscale Bioimaging: from Molecular Machines to
Networks of Excitable Cells" (MBExC), University of Göttingen, Germany\and
Johns Hopkins University, Baltimore, USA \and
Institute of AI for Health, Helmholtz Munich - German Research Center for Environmental Health, Neuherberg, Germany \and
Helmholtz AI, Helmholtz Munich - German Research Center for Environmental Health, Neuherberg, Germany\\
\email{jan.huisken@uni-goettingen.de, tingying.peng@helmholtz-muenchen.de}}

\maketitle              
\begin{abstract}
Light-sheet fluorescence microscopy (LSFM), a planar illumination technique that enables high-resolution imaging of samples, experiences “defocused” image quality caused by light scattering when photons propagate through thick tissues. To circumvent this issue, dual-view imaging is helpful. It allows various sections of the specimen to be scanned ideally by viewing the sample from opposing orientations. Recent image fusion approaches can then be applied to determine in-focus pixels by comparing image qualities of two views locally and thus yield spatially inconsistent focus measures due to their limited field-of-view. Here, we propose BigFUSE, a global context-aware image fuser that stabilizes image fusion in LSFM by considering the global impact of photon propagation in the specimen while determining focus-defocus based on local image qualities. Inspired by the image formation prior in dual-view LSFM, image fusion is considered as estimating a focus-defocus boundary using Bayes' Theorem, where (\textit{i}) the effect of light scattering onto focus measures is included within \textit{Likelihood}; and (\textit{ii}) the spatial consistency regarding focus-defocus is imposed in \textit{Prior}. The expectation-maximum algorithm is then adopted to estimate the focus-defocus boundary. Competitive experimental results show that BigFUSE is the first dual-view LSFM fuser that is able to exclude structured artifacts when fusing information, highlighting its abilities of automatic image fusion.

\end{abstract}

\section{Introduction}
Light-sheet fluorescence microscopy (LSFM), characterized by orthogonal illumination with respect to detection, provides higher imaging speeds than other light microscopies, e.g., confocal microscopy, via gentle optical sectioning \cite{power2017guide,reynaud2015guide,keller2008quantitative}, which makes it well-suited for whole-organism studies \cite{medeiros2015confocal}. At macroscopic scales, however, light scattering degrade image quality. It leads to images from deeper layers of the sample being of worse quality than from tissues close to the illumination source \cite{verveer2018restoration,krzic2012multiview}. To overcome the negative effect of photon propagation, dual-view LSFM is introduced, in which the sample is sequentially illuminated from opposing directions, and thus portions of the specimen with inferior quality in one view will be better in the other \cite{rubio2012wavelet} (Fig. \ref{overview}a). Thus, image fusion methods that combine information from opposite views into one volume are needed.

To realize dual-view LSFM fusion, recent pipelines adapt image fusers for natural image fusion to weigh between views by comparing the local clarity of images \cite{rubio2012wavelet,verveer2018restoration}. For example, one line of research estimates focus measures in a transformed domain, e.g., wavelet \cite{lewis2007pixel} or contourlet \cite{zhang2009multifocus}, such that details with various scales can be considered independently \cite{LIU202071}. However, the composite result often exhibits global artifacts \cite{li2013image}. Another line of studies conducts fusion in the image space, with pixel-level focus measures decided via local block-based representational engineerings such as multi-scale weighted multi-scale weighted gradient \cite{zhou2014multi} and SIFT \cite{liu2015multi}. Unfortunately, spatially inconsistent focus measures are commonly derived for LSFM, considering the sparse structures of the biological sample involved by the limited field-of-view (FOV). 

Apart from limited FOV, another obstacle that hinders the adoption of natural image fusion methods into dual-view LSFM is the inability to distinguish sample structures from structural artifacts \cite{AZAM2022105253}. For example, ghost artifacts, surrounding the sample as a result of scattered illumination light \cite{fahrbach2010microscopy}, can be observed, as it only appears in regions far from the light source after light travels through scattering tissues. Yet, when ghosts appear in one view, the same region in the opposite view would be background, i.e., no signal. Thus, ghosts will be erroneously transferred to the result by conventional fusion studies, as they are considered as owning richer information than its counterpart in the other view.

Here, we propose BigFUSE to realize spatially consistent image fusion and exclude ghost artifacts. Main contributions are summarized as follows:
\begin{itemize}[noitemsep,topsep=0pt]
\item[$\bullet$] BigFUSE is the first effort to think of dual-view LSFM fusion using Bayes, which maximizes the conditional probability of fused volume regarding image clarity, given the image formation prior of opposing illumination directions.
\item[$\bullet$] The overall focus measure along illumination is modeled as a joint consideration of both global light scattering and local neighboring image qualities in the contourlet domain, which, together with the smoothness of focus-defocus, can be maximized as \textit{Likelihood} and \textit{Prior} in Bayesian.
\item[$\bullet$] Aided by a reliable initialization, BigFUSE can be efficiently optimized by utilizing expectation-maximum (EM) algorithm.
\end{itemize}

\section{Methods}
\begin{figure}[h]
\centering
    \includegraphics[width=0.9\textwidth]{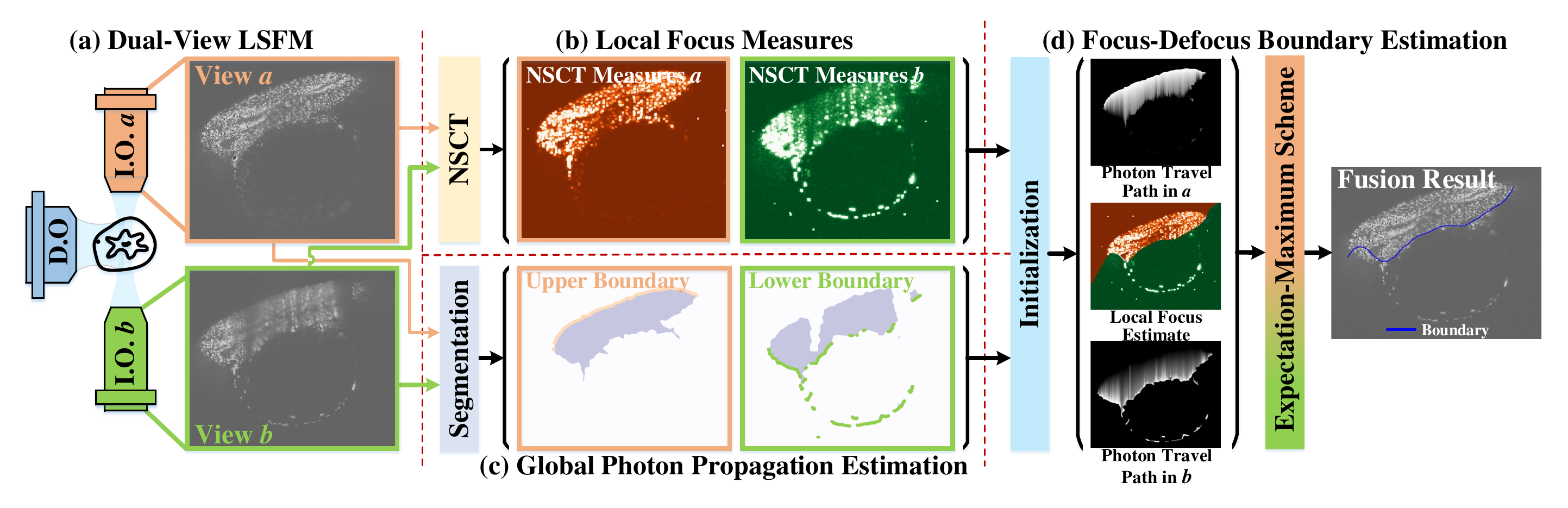}
    \caption{An overview of BigFUSE (see text for explanation)}
    \label{overview}
\end{figure}

An illustration of BigFUSE for dual-view LSFM fusion is given in Fig. \ref{overview}. First, pixel-level focus measures are derived for two opposing views separately, using nonsubsampled contourlet transform (NSCT) (Fig. \ref{overview}b). Pixel-wise photon propagation maps in tissue are then determined along light sheet via segmentation (Fig. \ref{overview}c). The overall focus measures are thus modeled as the inverse of integrated photon scattering along illumination conditioned on the focus-defocus change, i.e., \textit{Likelihood}, whereas the smoothness of focus-defocus is ensured via \textit{Prior}. Finally, the focus-defocus boundary is optimized via EM (Fig. \ref{overview}d).
\subsection{Revisiting Dual-View LSFM Fusion using Bayes}
BigFUSE first rethinks dual-view LSFM fusion from a Bayesian perspective, that is, the conditional probability of fused volume in terms of "in-focusness", is given on not only a pair of image inputs, but also prior knowledge that these two images are illiminated from opposing orientations respectively:
\begin{eqnarray}\label{eq:bayesian}
{\rm{Y}} = \mathop {{\rm{argmax}}}\limits_{\rm{Y}} p({\rm{Y}}|({{\rm{X}}^a},{{\rm{X}}^b}),{\cal P}) = \mathop {{\rm{argmax}}}\limits_{\rm{Y}} \{ p({\rm{Y}}|{\cal P})p(({{\rm{X}}^a},{{\rm{X}}^b})|{\rm{Y}},{\cal P})\}
\end{eqnarray}
where ${\rm{Y}} \in {{\mathbb{R}}^{M \times N}}$ is our predicted fusion with minimal light scattering effect. ${{\rm{X}}^a}$ and ${{\rm{X}}^b}$ are two image views illuminated by light source $a$ and $b$, respectively. We choose ${{\rm{Y}}}\in \{{{\rm{X}}^a},{{\rm{X}}^b}\}$ depending on their competitive image clarity at each pixel. Priors $\cal P$ denote our empirical favor of ${{\rm{X}}^a}$ against ${{\rm{X}}^b}$ at each pixel if photons travel through fewer scattering tissues from source \textit{a} than \textit{b}, and vice versa. Due to the non-positive light scattering effect along illumination path, there is only one focus-defocus change per column for dual-view LSFM in Fig. \ref{overview}a. Thus, fusion is equivalent to estimating a focus-defocus boundary $\omega$ defined as a function associating focus-defocus changes to column indexes:
\begin{eqnarray}\label{eq:beyes}
\omega  = \mathop {{\rm{argmax}}}\limits_\omega  p(\omega )p(({{\rm{X}}^a},{{\rm{X}}^b})|\omega)
\end{eqnarray}
which can be further reformulated by logarithm:
\begin{eqnarray}\label{eq:MAP}
\omega  &=& \mathop {{\rm{argmax}}}\limits_\omega  p(\omega )p(({{\rm{X}}^a},{{\rm{X}}^b})|\omega)
= \mathop {{\rm{argmax}}}\limits_{\omega  \in \{ 1,2, \ldots ,N\} } {\mkern 1mu} {\rm{log}}(p(\omega )\prod\nolimits_{i = 1}^N {p(({\rm{X}}_{:,i}^a,{\rm{X}}_{:,i}^b)|{\omega _i})} ) \nonumber\\ &=& \mathop {{\rm{argmax}}}\limits_{\omega  \in \{ 1,2, \ldots ,N\} } \left\{ {{\rm{log}}(p(\omega )) + \sum\nolimits_{i = 1}^N {{\rm{log}}(p(({\rm{X}}_{:,i}^a,{\rm{X}}_{:,i}^b)|{\omega _i}))} } \right\}{\mkern 1mu}
\end{eqnarray}
where ${\omega _i}$ denotes the focus-defocus changeover at \textit{i}-th column, ${{\rm{X}}_{:,i}}$ is the \textit{i}-th column of ${\rm{X}}$. Next, estimating $\omega$ is decomposed into: (\textit{i}) define the column-wise image clarity, i.e., \textit{log-likelihood} ${\rm{log}}({p(({\rm{X}}_{:,i}^a,{\rm{X}}_{:,i}^b)|{\omega _i})})$; (\textit{ii}) consider the belief on a spatially smooth focus-defocus boundary, namely \textit{log-prior} ${\rm{log}}(p(\omega ))$.

\subsection{Image Clarity Characterization with Image Formation Prior}
In LSFM, log-likelihood ${\rm{log}}({p(({\rm{X}}_{:,i}^a,{\rm{X}}_{:,i}^b)|{\omega _i})})$ can be interpreted as the probability of observing $({\rm{X}}_{:,i}^a,{\rm{X}}_{:,i}^b)$ given the hypothesis that the focus-defocus change is determined as $\omega_i$ in the \textit{i}-th column:
\begin{eqnarray}\label{eq:Log-likelihood}
\mathop {{\rm{argmax}}}\limits_{{\omega _i}} \left\{ {{\rm{log}}(p(({\rm{X}}_{:,i}^a,{\rm{X}}_{:,i}^b)|{\omega _i}))} \right\} = \mathop {{\rm{argmax}}}\limits_{{\omega _i}} \left\{ {c({\rm{X}}_{1:{\omega _i},i}^a \oplus {\rm{X}}_{{\omega _i} + 1:M,i}^b)} \right\}
\end{eqnarray}
where $\oplus$ is a concatenation, $c(\bullet)$ is the columne-wise image clarity to be defined.

\subsubsection{Estimating Pixel-Level Image Clarity in NSCT.} To define $c(\bullet)$, BigFUSE first uses NSCT, a shift-invariant image representation technique, to estimate pixel-level focus measures by characterizing salient image structures \cite{zhang2009multifocus}. Specifically, NSCT coefficients ${\rm{S}}^a$ and ${\rm{S}}^b$ are derived for two opposing LSFM views, where ${\rm{S}} = \{ {{\rm{S}}_{{i_0}}},{{\rm{S}}_{i,l}}|(1 \le i \le {i_0},1 \le l \le {2^{{l_i}}})\}$, ${\rm{S}}_{{i_0}}$ is the lowpass coefficient at the coarsest scale, ${\rm{S}}_{i, l}$ is the bandpass directional coefficient at \textit{i}-th scale and \textit{l}-th direction. Local image clarity is then projected from ${\rm{S}}^a$ and ${\rm{S}}^b$ \cite{zhang2009multifocus}:
\begin{eqnarray}\label{eq:Principle}
{{\rm{F}}_{i,l}} = {{\rm{R}}_{i,l}} \times {\rm{D}}{\sigma _i} = \frac{{\left| {{{\rm{S}}_{i,l}}} \right|}}{{{{{\rm{\bar S}}}_{{i_0}}}}} \times \sqrt {\frac{1}{{{2^{{l_i}}}}}\sum\nolimits_{r = 1}^{{2^{{l_i}}}} {{{[\left| {{S_{i,r}}} \right| - {{\bar S}_i}]}^2}} } ,\hspace{0.5em} {\rm{   }}{{\rm{\bar S}}_i} = \frac{1}{{{2^{{l_i}}}}}\sum\nolimits_{r = 1}^{{2^{{l_i}}}} {\left| {{{\rm{S}}_{i,r}}} \right|}
\end{eqnarray}
where ${\rm{R}}_{i,l}$ is local directional band-limited image contrast and ${{\rm{\bar S}}}_{{i_0}}$ is the smoothed image baseline, whereas $ {\rm{D}}{\sigma _i}$ highlights image features that are distributed only on a few directions, which is helpful to exclude noise \cite{zhang2009multifocus}. As a result, pixel-level image clarity ${\rm{F}}=\sum\nolimits_{j=1}^{{{j}_{0}}}{\sum\nolimits_{l=1}^{2\hat{\ }{{l}_{j}}}{{{{\rm{F}}}_{j,l}}}}$ is quantified for respective LSFM view. 

\subsubsection{Reweighting Image Clarity Measures by Photon Traveling Path.} Pixel-independent focus measures may be adversely sensitive to noise, due to the limited receptive field when characterizing local image clarities. Thus, BigFUSE proposes to integrate pixel-independent image clarity measures along columns by taking into consideration the photon propagation in depth. Specifically, given a pair of pixels $(\text{X}_{m,n}^{a},\text{X}_{m,n}^{b})$, $\text{X}_{m,n}^{a}$ is empirically more in-focus than $\text{X}_{m,n}^{a}$, if photons travel through fewer light-scattering tissues from illumination objective \textit{a} than from \textit{b} to get to position (\textit{m}, \textit{n}), and vice versa. Therefore, BigFUSE defines column-level image clarity measures as:
\begin{eqnarray}\label{eq:coclarity}
c({\rm{X}}_{1:{\omega _i},i}^a \oplus {\rm{X}}_{{\omega _i} + 1:M,i}^b) = \sum\nolimits_{j = 1}^{{\omega _i}} {{{\rm{A}}_{j,i}}{\rm{F}}_{j,i}^a}  + \sum\nolimits_{j = {\omega _i} + 1}^M {{{\rm{A}}_{j,i}}{\rm{F}}_{j,i}^b}
\end{eqnarray}
where ${{\text{A}}_{:,i}}$ is to model the image deterioration due to  light scattering. To visualize photon traveling path, BigFUSE uses OTSU thresholding for foreground segmentation (followed by AlphaShape to generalize bounding polygons), and thus obtains sample boundary, i.e., incident points of light sheet, which we refer to as $p^u$ and $p^l$ for opposing views \textit{a} and \textit{b} respectively. Since the derivative of ${{\text{A}}_{:,i}}$ implicitly refers to the spatially varying index of refraction within the sample, which is nearly impossible to accurately measure from the physics perspective, we model it using a piecewise linear model, without loss of generality:
\begin{eqnarray}\label{eq:linear}
{{\rm{A}}_{j,i}} = \left\{ {\begin{array}{*{20}{l}}
{1 - 0.5 \times \left| {j - {p^u}} \right|/{\omega _i},}&{j < {\omega _i}}\\
{0.5 + 0.5 \times \left| {j - {p^l}} \right|/(M - {\omega _i}),}&{j \ge {\omega _i}}\\
{0,}&{{\rm{otherwise}}}
\end{array}} \right.
\end{eqnarray}
As a result, ${\rm{log}}({p(({\rm{X}}_{:,i}^a,{\rm{X}}_{:,i}^b)|{\omega _i})})$ is obtained as summed pixel-level image clarity measures with integral factors conditioned on photon propagation in depth.

\subsection{Least Squares Smoothness of Focus-Defocus Boundary}
With log-likelihood ${\rm{log}}({p(({\rm{X}}_{:,i}^a,{\rm{X}}_{:,i}^b)|{\omega _i})})$ considering the focus-defocus consistency along illuminations using image formation prior in LSFM, BigFUSE then ensures consistency across columns in $p(\omega)$. Specifically, the smoothness of $\omega$ is characterized as a window-based polynomial fitness using linear least squares:
\begin{eqnarray}\label{eq:prior}
{\rm{log}}(p(\omega )) = \sum\nolimits_{i = 1}^N {{\rm{log}}(p({\omega _i}))}  = \sum\nolimits_{i = 1}^N {{{\left\| {{\omega _{i - s:i + s}} - {{\hat \upsilon }_i}} \right\|}^2}}
\end{eqnarray}
where ${\hat \upsilon _i} = {c_i}{{\rm{\Omega}} _i} = [{c_{i,0}},{c_{i,1}}, \ldots ,{c_{i,Q}}][\boldsymbol{\omega} _{i - s}^{\rm{T}},\boldsymbol{\omega} _{i - s + 1}^{\rm{T}}, \ldots ,\boldsymbol{\omega} _{i + s}^{\rm{T}}]$, ${\boldsymbol{\omega} _i} = [\omega _i^Q,\omega _i^{Q - 1}, \ldots ,1]$, ${c_i}$ is the parameters to be estimated, the sliding window is with a size of $2s+1$.

\subsection{Focus-Defocus Boundary Inference via EM}
Finally, in order to estimate the $\omega$ together with the fitting parameter $c$, BigFUSE reformulates the the posterior distribution in Eq. (\ref{eq:beyes}) as follows:
\begin{equation}\label{eq:energyFuction}
\begin{aligned}
\{ \hat \omega ,\hat c\}  = \mathop {{\rm{argmax}}}\limits_{\omega ,c} {\mkern 1mu} \left\{ \begin{array}{l}
\sum\nolimits_{j = 1}^N {\left\{ {\sum\nolimits_{i = 1}^{{\omega _j}} {{{\rm{A}}_{i,j}}{\rm{F}}_{i,j}^a}  + \sum\nolimits_{i = {\omega _j} + 1}^M {{{\rm{A}}_{i,j}}{\rm{F}}_{i,j}^b} } \right\}} \\
 + \lambda \{ \sum\nolimits_{i = 1}^N {{{\left\| {{\omega _{i - s:i + s}} - {{\hat \upsilon }_i}} \right\|}^2}}\} \end{array} \right\}
\end{aligned}
\end{equation}
where $\lambda$ is the trade-off parameter. Here, BigFUSE alternates the estimations of $\omega$, and $c$, and iterates until the method converges. Specifically, given $c^{(n)}$ for the \textit{n}-th iteration, ${{\omega }_{i}^{(n+1)}}$ is estimated by maximizing (E-step):
\begin{equation}\label{eq:part1}
\begin{aligned}
\omega _i^{(n + 1)} = \mathop {{\rm{argmax}}}\limits_{{\omega _i}} \left\{ {\sum\nolimits_{j = 1}^{{\omega _i}} {{{\rm{A}}_{j,i}}{\rm{F}}_{j,i}^a}  + \sum\nolimits_{j = {\omega _i} + 1}^M {{{\rm{A}}_{j,i}}{\rm{F}}_{j,i}^b}  + \lambda {{\left\| {{\omega _{i - s:i + s}} - {{\hat \upsilon }_i}} \right\|}^2}} \right\}
\end{aligned}
\end{equation}
which can be solved by iterating over $\{i|1\le i\le M\}$. BigFUSE then updates $c_{i}^{(n+1)}$ based on least squares estimation:
\begin{equation}\label{eq:part2}
\begin{aligned}
c_i^{(n + 1)} = {({\rm{\Omega} ^{\rm{T}}}\rm{\Omega} )^{ - 1}}{\rm{\Omega} ^{\rm{T}}}{\omega _{i - s:i + s}}
\end{aligned}
\end{equation}
Additionally, ${\rm{A}}^{n+1}$ is updated based on Eq. (\ref{eq:linear}) subject to $\omega^{(n+1)}$ (M-step). BigFUSE proposes to initialize $\omega$ based on ${\rm{F}}^a$ and ${\rm{F}}^b$:
\begin{equation}\label{eq:initial}
\begin{aligned}
\omega _i^{(0)} = {p^u} + |{\rm{\Phi}} |,\quad{\rm{\Phi}}  = \{ j|{\rm{F}}_{j,i}^a > {\rm{F}}_{j,i}^b,{p^u} \le j \le {p^l}\}
\end{aligned}
\end{equation}
where $|{\rm{\Phi}} |$ denotes the total number of elements in ${\rm{\Phi}}$. 

\subsection{Competitive Methods}
We compare BigFUSE to four baseline methods: (\textit{i}) DWT \cite{rubio2012wavelet}: a multi-resolution image fusion method using discrete wavelet transform (DWT); (\textit{ii}) NSCT \cite{zhang2009multifocus}: another multi-scale image fuser but in the NSCT domain; (\textit{iii}) dSIFT \cite{liu2015multi}: a dense SIFT-based focus estimator in the image space; (\textit{iv}) BF \cite{zhang2017boundary}: a focus-defocus boundary detection method that considers region consistency in focus; and two BigFUSE variations: (\textit{v}) $\mathcal{S}(\bullet)$: built by disabling smooth constraint; (\textit{vi}) $\mathcal{P}(\bullet)$: formulated by replacing the weighted summation of pixel-level clarity measures for overall characterization, by a simple average.

To access the blind image fusion performance, we adopt three fusion quality metrics, $Q_{mi}$ \cite{hossny2008comments}, $Q_{g}$ \cite{petrovic2005objective} and $Q_{s}$ \cite{piella2003new}. Specifically, $Q_{mi}$, $Q_{g}$ and $Q_{s}$ use mutual information, gradient or image quality index to quantify how well the information or features of the inputs are transferred to the result, respectively. In the simulation studies where ground truth is available, mean square error (EMSE) and structural similarity index (SSIM) are used for quantification.
\begin{table}[tp]
   \centering
   \begin{threeparttable}[b]
   \caption{BigFUSE achieves best quantitative results on synthetic blur.}
   \label{tab:test2}
   \centering
\begin{tabular}{c|c|c|c|c|c|c|c}
\toprule[2pt]
\rule{0pt}{0pt}
     & \makecell[c]{DWT \cite{rubio2012wavelet}}   
     & \makecell[c]{NSCT \cite{zhang2009multifocus}}          
     & \makecell[c]{dSIFT \cite{liu2015multi}} 
     & \makecell[c]{BF \cite{zhang2017boundary}} 
     & \makecell[c]{$\mathcal{S}( \bullet )$}
     & \makecell[c]{$\mathcal{P}( \bullet )$}
     & \textbf{BigFUSE}       \\[0pt]
\hline
\rule{0pt}{15pt}
\makecell[c]{EMSE \\ ($\times 10^{-5}$)} & \makecell[c]{6.81 \\ $\pm$0.72} & \makecell[c]{4.66 \\ $\pm$0.56} & \makecell[c]{5.17 \\ $\pm$0.72}    & \makecell[c]{1.55 \\ $\pm$0.25} & \makecell[c]{1.43 \\ $\pm$0.34}    & \makecell[c]{0.96 \\ $\pm$0.08}    & \makecell[c]{\textbf{0.94} \\ $\pm$\textbf{0.09}} \\[3pt]
\rule{0pt}{15pt}
SSIM & \makecell[c]{0.974 \\ $\pm$0.02}  & \makecell[c]{0.996 \\ $\pm$0.03}  & \makecell[c]{0.93 \\ $\pm$0.01}     & \makecell[c]{0.994 \\ $\pm$0.03}  & \makecell[c]{0.993 \\ $\pm$0.02}
  & \makecell[c]{0.993 \\ $\pm$0.02}.    & \makecell[c]{\textbf{0.998} \\ $\pm$\textbf{0.01}} \\[0pt]
\bottomrule[2pt]
\end{tabular}
  \end{threeparttable}
\end{table}
\begin{figure}[tp]
\centering
    \includegraphics[width=0.9\textwidth]{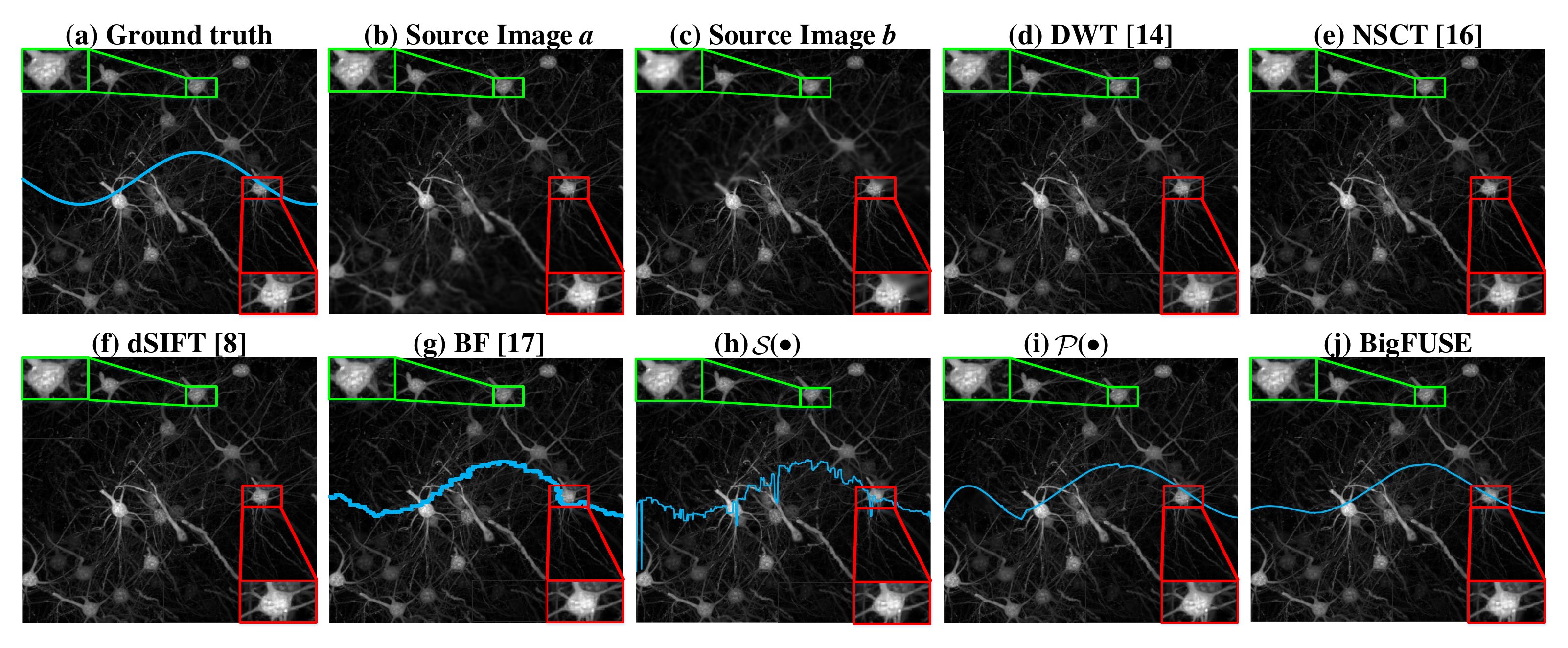}
    \caption{Visualization of fusion quality with respect to ground-truth. Focus-defocus boundary is given in blue curve, if applicable.}
    \label{fakeresult}
\end{figure}

\section{Results and Discussion}
\subsection{Evaluation on LSFM Images with Synthetic Blur}
We first evaluate BigFUSE in fusing dual-view LSFM blurred by simulation. Here, we blur a mouse brain sample collected in \cite{dean2022isotropic} with spatially varying Gaussian filter for thirty times, which is chemically well-cleared and thus provides an almost optimal ground truth for simulation, perform image fusion, and compare the results to the ground truth. BigFUSE achieves the best EMSE and SSIM, statistically surpassing other approaches (Table. \ref{tab:test2}, $p < 0.05$ using Wilcoxon signed-rank test). Only BigFUSE and BF realize information fusing without damaging original images with top two EMSE. In comparison, DWT, NSCT and dSIFT could distort the original signal when fusing (green box in Fig. \ref{fakeresult}).
\begin{figure}[tp]
\centering
    \includegraphics[width=1\textwidth]{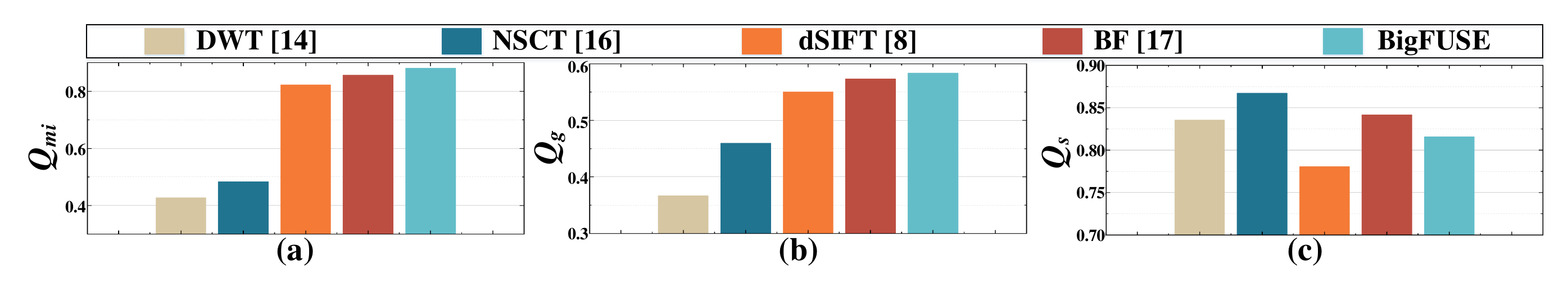}
    \caption{Comparison between BigFUSE and baselines on $Q_{mi}$, $Q_{g}$ and $Q_{s}$.}
    \label{realresultnumber}
\end{figure}
\begin{figure}[tp]
\centering
    \includegraphics[width=1\textwidth]{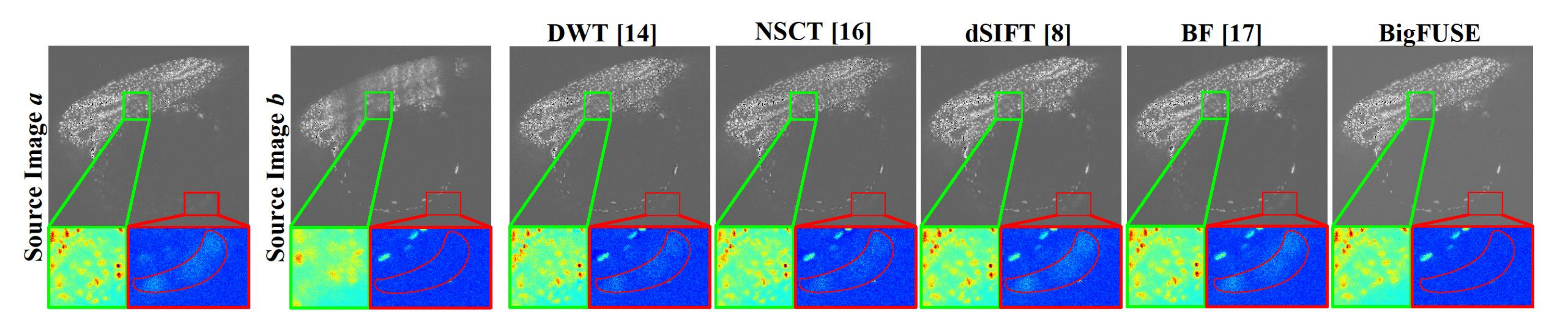}
    \caption{Visualization of fusing a zebrafish embryo.}
    \label{realresult}
\end{figure}
\begin{figure}[tp]
\centering
    \includegraphics[width=1\textwidth]{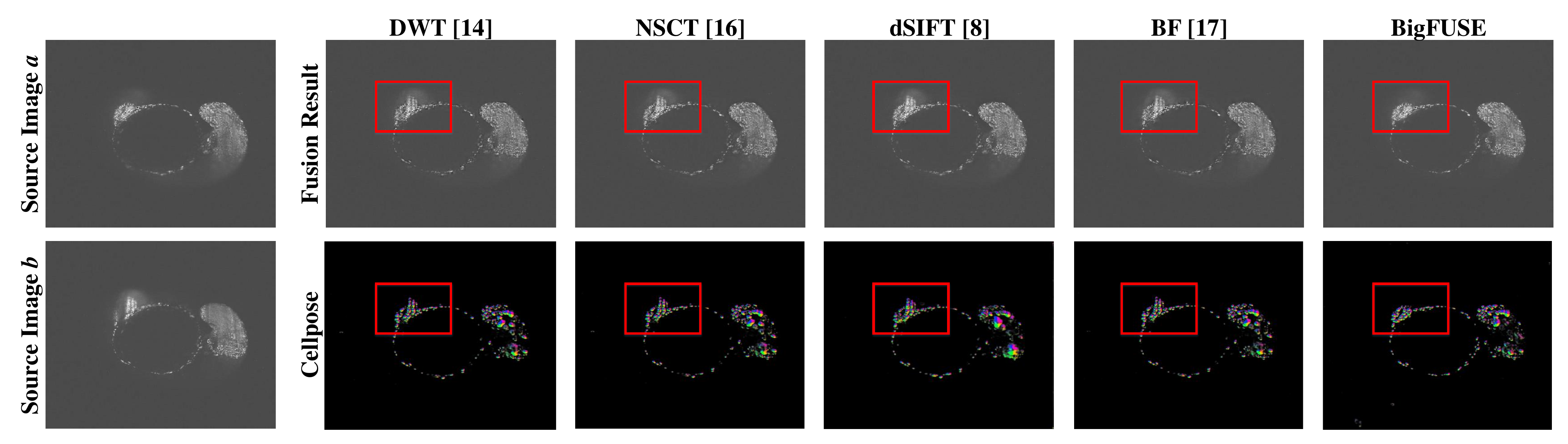}
    \caption{Visualization of Cellpose result.}
    \label{seg}
\end{figure}
\subsection{Evaluation on LSFM Images with Real Blur}
BigFUSE is then evaluated against baseline methods on real dual-view LSFM. A large sample volume, zebrafish embryo ($282 \times 2048 \times 2048$ for each view), is imaged using a Flamingo Light Sheet Microscope. BigFUSE takes roughly nine minutes to process this zebrafish embryo, using a T4 GPU with 25 GB system RAM and 15 GB GPU RAM. In Fig. \ref{realresult}, inconsistent boundary is detected by BF, while methods like DWT and NSCT generate structures that do not exist in either input. Moreover, only BigFUSE can exclude ghost artifact from the result (red box), as BigFUSE is the only pipeline that considers image formation prior. Additionally, we demonstrate the impact of bigFUSE on a specific downstream task in Fig. \ref{seg}, i.e., segmentation by Cellpose. Only the fusion result provided by bigFUSE allows reasonable predicted cell pose, given that ghosting artifacts are excluded dramatically. This explains why the $Q_s$ in Fig. \ref{realresultnumber} is suboptimal, since BigFUSE do not allow for the transmission of structural ghosts to the output.

\section{Conclusion}
In this paper, we propose BigFUSE, a image fusion pipline with image formation prior. Specifically, image fusion in dual-view LSFM is revisited as inferring a focus-defocus boundary using Bayes, which is essential to exclude ghost artifacts. Furthermore, focus measures are determined based on not only pure image representational engineering in NSCT domain, but also the empirical effects of photon propagation in depth embeded in the opposite illumination directions in dual-view LSFM. 
BigFUSE can be efficiently optimized using EM. Both qualitative and quantitative evaluations show that BigFUSE surpasses other state-of-the-art LSFM fusers by a large margin. BigFUSE will be made accessible.

\subsubsection{Acknowledgements} 
Y.L. is supported by the China Scholarship Council (No. 202106020050).
J.H. is funded by the Alexander von Humboldt Foundation in the framework of the Alexander von Humboldt Professorship endowed by the German Federal Ministry of Education and Research and funded by the Deutsche Forschungsgemeinschaft (DFG, German Research Foundation) under Germany’s Excellence Strategy - EXC 2067/1- 390729940.
\bibliographystyle{splncs04}
\bibliography{03mybibliography}
\end{document}